\documentclass{INTERSPEECH2023}

\interspeechcameraready

\title{Mandarin Electrolaryngeal Speech Voice Conversion \\using Cross-domain Features}
\name{
      Hsin-Hao Chen$^{1,2}$, Yung-Lun Chien$^{1,2}$, Ming-Chi Yen$^2$, Shu-Wei Tsai$^3$, \\Yu Tsao$^2$, Tai-shih Chi$^1$, and Hsin-Min Wang$^2$
}
\address{
 $^1$ National Yang Ming Chiao Tung University,
$^2$ Academia Sinica,\\
$^3$ National Cheng Kung University Hospital }
\email{123ggg3304@gmail.com,ajul1230@gmail.com,ymchiqq@iis.sinica.edu.tw,tsaisuwei@gmail.com,\\yu.tsao@citi.sinica.edu.tw,tschi@mail.nctu.edu.tw, whm@iis.sinica.edu.tw}

\begin{document}

\maketitle
 
\begin{abstract}
Patients who have had their entire larynx removed, including the vocal folds, owing to throat cancer may experience difficulties in speaking. In such cases, electrolarynx devices are often prescribed to produce speech, which is commonly referred to as electrolaryngeal speech (EL speech). However, the quality and intelligibility of EL speech are poor.
To address this problem, EL voice conversion (ELVC) is a method used to improve the intelligibility and quality of EL speech. In this paper, we propose a novel ELVC system that incorporates cross-domain features, specifically spectral features and self-supervised learning (SSL) embeddings.
The experimental results show that applying cross-domain features can notably improve the conversion performance for the ELVC task compared with utilizing only traditional spectral features.  
\end{abstract}

\section{Introduction}
\label{sec:intro}
Some patients with laryngeal diseases, such as laryngeal cancer, may need to undergo laryngectomy, which is a surgical procedure that involves removing the larynx, including the vocal folds. Without the vibration of the vocal folds, these patients lose  their ability to generate excitation signals and cannot produce speech normally. Electrolarynx (EL) is a medical device used to restore the ability to speak in these patients. Specifically, the electrolarynx generates surrogate excitation signals that enable patients to produce speech. However, the sound quality and intelligibility of the speech produced by the electrolarynx are often poor and accompanied by constant mechanical noise, and consequently, the speech does not resemble a natural human voice.

Voice conversion (VC)\cite{kameoka2018stargan,qian2019autovc,sisman2020overview} techniques have been widely used to overcome this problem to convert electrolaryngeal (EL) speech  to natural (NL) speech without altering the underlying content. This task is commonly referred to as EL speech voice conversion (ELVC) \cite{nakamura2012speaking,li2017mandarin,malathi2018enhancement,kobayashi2018,kobayashi2021,dinh2020increasing,yen2021mandarin}. Existing ELVC approaches can be classified into two categories: sequence-to-sequence and frame-based approaches.


The implementation of a frame-based ELVC system typically involves three steps: first, extracting features from both EL and NL speech; second, converting the features of the EL speech to those of the target NL speech using a conversion model; and finally, generating speech waveforms from the converted features, which is often performed by a vocoder. Traditional frame-based ELVC techniques, such as those used in \cite{nakamura2012speaking, li2017mandarin,malathi2018enhancement}, employ Gaussian mixture models to build the conversion model. Recently, deep neural networks have been widely used to build the conversion model \cite{kobayashi2018,kobayashi2021,dinh2020increasing}.


In addition to frame-based ELVC systems, previous studies have proposed implementing ELVC systems using sequence-to-sequence (seq2seq) models. A seq2seq ELVC system can effectively avoid frame-alignment difficulties that can occur in frame-based ELVC systems \cite{yen2021mandarin}. However, a relatively large amount of computation is required during training and conversion, which makes such systems challenging to implement in real-world scenarios.


\begin{figure*}[t]
    \centering
    \includegraphics[width=\linewidth]{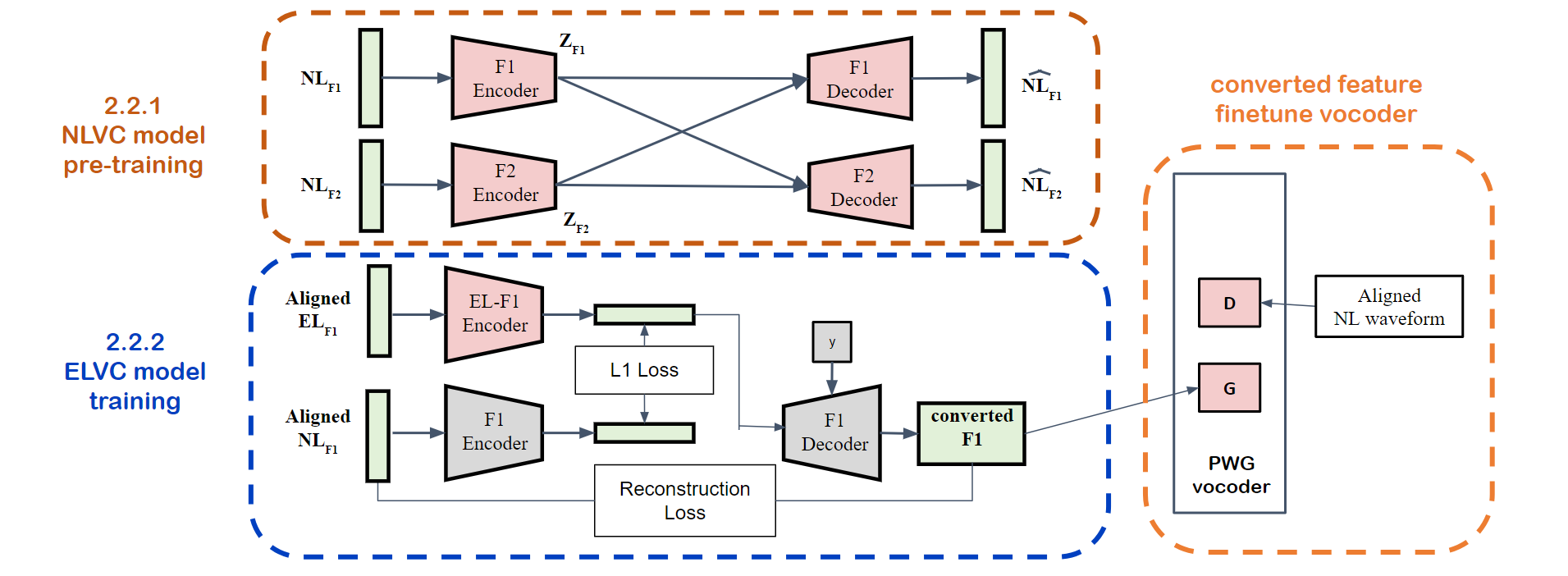}
    \caption{The overall architecture of the proposed ELVC system. F1 and F2 represent features 1 and 2, respectively, and y represents the speaker identity. During NLVC pre-training, the encoder and decoder (pink parts) are tuned based on NL speech. During ELVC training, the gray part is fixed and the pink part is tuned based on aligned NL and EL speech. $Z_{F1}$ and $Z_{F2} $ denote the latent representations generated from the F1 and F2 encoders, respectively, and the G and D components of the PWG vocoder refer to the generator and discriminator of the GAN-based network.}
    \label{fig:all}
\end{figure*}

Self-supervised learning (SSL) has become popular in recent years. It is a promising alternative to traditional supervised learning, which requires large amounts of labeled data, which can be both time-consuming and labor-intensive. By contrast, SSL can train high-performance models without the need for labeled data, thereby overcoming the aforementioned limitations of supervised learning. The representations obtained from SSL (referred to as SSL representations) have proven to be highly effective in various speech-related tasks, as demonstrated in several studies \cite{baevski2020wav2vec,hsu2021hubert,chen2022wavlm}. In the context of the VC task, SSL representations have also been incorporated and shown promising results in one-to-one, any-to-one, and any-to-any modes \cite{lin2021s2vc,huang2021any}.


In \cite{hernandez2022cross}, the use of SSL representations improved speech recognition for patients with dysarthria, which is a type of impaired speech. However, dysarthric  speech and EL speech have distinct characteristics. The dysarthric speech dataset used in \cite{hernandez2022cross} includes speech signals from patients with cerebral palsy and Parkinson’s disease, which cause motor control dysfunction, primarily affecting pronunciation. In contrast, EL speech is generated from surrogate excitation signals, which are significantly different from the real excitation signals generated from the lungs. In \cite{violeta2022investigating}, the authors investigated an SSL pre-trained  model for pathological speech, focusing on fine-tuning the pre-trained  model for speech recognition.



In this paper, we propose a novel frame-based Mandarin ELVC system that utilizes both traditional spectral features and SSL representations (or called SSL features) extracted using WavLM \cite{chen2022wavlm}. The ELVC system comprises three steps, and we employ SSL and cross-domain features in all these steps. First, we used the SSL features to implement a dynamic time warping (DTW) algorithm to align the EL and NL speech signals. Subsequently, we employed the cross-domain variational autoencoder (CDVAE) \cite{huang2018voice} as a conversion model to convert EL speech to NL speech. Finally, we trained a vocoder using cross-domain features to generate speech audio from the converted features. In summary, the contributions of this study are as follows: (1) confirming the effectiveness of the SSL features for temporal alignment, (2) confirming the effectiveness of the cross-domain features, including the SSL features, for the ELVC task, and (3) training vocoders based on the SSL  and cross-domain features. To the best of our knowledge, this is the first study to investigate this topic with promising results. We also demonstrate the effectiveness of the proposed method through subjective and objective evaluations.

\section{Methods}
\label{sec:Methods}
In this section, we introduce our overall architecture (see Fig.  1). The frame-based ELVC system is implemented in three steps: feature extraction, feature conversion, and audio generation. Before generating the audio file, we used the converted features to fine-tune the vocoder to obtain better results.
\subsection{Feature extraction}

We used cross-domain features in two parts: (1) traditional features: mel spectrum (Mel), mel cepstral coefficients (MCC), and STRAIGHT spectra (SP); (2) SSL features, which refers to the embeddings extracted by the pre-trained WavLM model \cite{WavLM_links} from the input waveform.

\subsection{Feature conversion}
Owing to the scarcity of EL data, we propose a two-stage feature conversion approach. In the first stage, we trained an NL speech conversion (NLVC) model as a pre-trained model. In the second stage, we fine-tuned the pre-trained NLVC model using EL speech data to obtain an ELVC model. This approach leverages the abundance of NL speech data to improve the conversion performance of the limited EL speech data.

\subsubsection{Stage1. NLVC model training}
We trained the NLVC model using utterances from 18 speakers. The model adopts a CDVAE  architecture \cite{huang2018voice}. This model benefits from the simultaneous use of multiple spectral features. In this study, we tested two sets of cross-domain features: Mel+SP and Mel+SSL. In the CDVAE model, the encoder uses cross-domain features as input and maps them to a latent space. The speaker identity was then concatenated with the latent representation before it was passed to the decoder to generate the predicted features.

\subsubsection{Stage2. Speech alignment and ELVC model training}
Before training the ELVC model, we aligned the EL and NL speech. We first hand-labeled word boundaries to split the entire sentence into word-by-word segments and then used the DTW algorithm to align the NL and EL segments. The DTW algorithm minimizes the distance between the NL and EL segments based on the mean square error (MSE). We implemented the DTW algorithm with different features, such as the Mel, MCC, and SSL features, as input, and the corresponding experimental results are presented in Section 4. With the aligned EL and NL data, we fed the NL features to the fixed NL encoder and the EL features to the EL encoder to train the VC model to minimize the difference between the two latent representations, where the L1 loss was used as the objective function to measure the difference. We then appended the speaker code to the latent representations and fed them to a decoder that was well-trained in Stage 1 to obtain the converted features. We derived another loss (reconstruction loss) to measure the difference between the converted and  NL speech features.

\subsection{Audio generation}
This section describes the process of generating audio files from converted features, which involves vocoder training and fine-tuning using cross-domain features. In this study, we used the Parallel WaveGAN (PWG) \cite{yamamoto2020parallel} vocoder because of its ability to produce high-quality audio and its efficient training and synthesis processes. 

We trained several vocoders using the Mel, SSL, and cross-domain features and compared the quality of the audio produced by them. Our experimental results for the NLVC task indicate that the PWG trained with cross-domain features is capable of producing high-quality audio. In contrast, for ELVC, the PWG trained with SSL features produces audio with better intelligibility. The above results demonstrate the effectiveness of incorporating SSL features into the PWG vocoder for both NLVC and ELVC tasks. Furthermore, we attempted to fine-tune the PWG model based on the converted features to enhance its performance specifically for the ELVC task.

\section{Experiments}
\subsection{Experimental setup}
We recruited a doctor (medical) to prepare the EL speech data by imitating a patient reading the Taiwanese Mandarin Noise Test (TMHINT) phonetically balanced script \cite{huang2005development} using an electrolarynx device. The EL data contained 320 utterances, of which 240 utterances were used for training. For the NL data, 18 speakers were recruited to read prompts in the TMHINT script. All the speech signals were recorded at a sampling rate of 16 kHz. For feature extraction, we used the World vocoder \cite{morise2016world} to extract 513-dimensional SP  and 24-dimensional MCCs. The frame size and number of hops were 1024 and 256, and the time shift was 20 ms. We also used this setting to extract the 80-dimensional mel spectrogram (Mel).  For the SSL feature, we used a pre-trained WavLM model with a frame/hop size of 400/320, resulting in a time shift of 25 ms, and each SSL feature had 768 dimensions. When using SSL features, we adopted the same frame size and hop size (400/320) to extract Mel features for consistency.

\subsubsection{CDVAE Model structure and parameters}
The CDVAE model consisted of two encoders and two decoders, each consisting of five convolutional neural network (CNN) layers. Assuming that the cross-domain features have N dimensions, the input dimension of the CNN layer in the encoder was (N, 1024, 512, 256, 128), and the output dimension of the decoder was (128, 256, 512, 1024, N). All the CNN layers in the model shared the same stride and kernel size of 1 and 5, respectively. The optimizer used in this model was RAdam (Rectified Adam), The batch size was set to 16, and the learning rate was set to 0.0001.

\subsubsection{Evaluation metrics}
Three objective metrics were employed to evaluate the proposed system: (1) mel-cepstral distortion (MCD) in dB, which measures spectral distortions; (2) fundamental frequency root mean square error (F0 RMSE) in Hz, which measures the accuracy of F0 information; and (3) fundamental frequency correlation coefficient (F0 CORR) in Hz, which measures the correlation of F0 features.

In addition, we conducted a subjective listening test in which participants rated the intelligibility and quality (cleanliness) of the converted audio files on a five-point scale, where 1 denotes the worst score and 5 denotes the best score. The listening test involved 12 untrained, normal hearing participants,  of whom 8 were male and 4 were female. The average age of the participants was 24 years old. For each test sample, participants were unaware which ELVC system was used to generate it. We selected 20 speech utterances from each ELVC system and asked subjects to rate them in terms of intelligibility and quality.

\subsection{Experimental results}
\subsubsection{DTW using different features}
In our initial experiments, we tested the performance of DTW using different features, including Mel, MCC, and WavLM. The CDVAE model for NLVC was trained using Mel as the NL$_{\rm{F1}}$ features and SP as the NL$_{\rm{F2}}$ features (see Fig. 1). The ELVC model was trained using Mel as the NL$_{\rm{F1}}$ and EL$_{\rm{F1}}$ features. The PWG vocoder was trained to generate speech audio from Mel. The results are shown in Table 1. The  SSL features yielded the best results (lowest F0 RMSE and highest F0 CORR), indicating that they are a better choice than the other features. We also tested DTW performance using cross-domain features, but no further improvements were observed. Therefore, we used the SSL features as input to the DTW algorithm in the following experiments.

\subsubsection{Vocoder using cross-domain features}
This section provides a performance comparison of different features for training the vocoder using NL speech from 18 speakers. Table 2 shows that the Mel features outperformed the SSL features, but the cross-domain features (SSL+Mel) yielded the best performance among the three systems. These results demonstrate the significant advantages of using cross-domain features to build an NL vocoder.

\begin{table}[h]
\caption{ELVC results when using different features in DTW.}
\centering
\begin{tabular}{c|ccc}
\text{DTW } & \text{MCD}  & \text{F0 RMSE} & \text{F0 CORR} \\ \hline
Mel                  & \textbf{8.43} & 35.48             & 0.078             \\
MCC                  & 8.46          & 35.41             & 0.100             \\
SSL                & 8.46          & \textbf{34.22}    & \textbf{0.131}   
\end{tabular}

\label{tab:my-table}
\end{table}

\begin{table}[h]
\caption{NL speech self-reconstruction results using vocoders with different features.}
\centering
\begin{tabular}{c|ccc}
\text{Vocoder}  & \text{MCD} & \text{F0 RMSE} & \text{F0 CORR }    \\ \hline
Mel    & \textbf{3.67} & 39.63          & 0.223          \\
SSL & 5.40          & 42.50          & 0.124          \\
SSL+Mel & 3.97          & \textbf{30.98} & \textbf{0.384}
\end{tabular}
\label{tab:my-table}
\end{table}

\begin{table}[h]
\caption{Overall results of ELVC, where the features used in CDVAE are denoted as F1+F2, and FT denotes the model after a further fine-tuning process.}
\centering
\begin{tabular}{cc|ccc}
 CDVAE & Vocoder                & MCD           & F0 RMSE        & F0 CORR        \\ \hline
Mel+SP                          & Mel            & 8.46          & \textbf{34.22} & 0.131          \\
Mel+SSL                       & Mel           & 7.68          & 42.01          & 0.041          \\
SSL+Mel                       & SSL         & 7.03          & 51.21          & 0.041          \\
SSL+Mel                       & SSL(FT)     & \textbf{6.90} & 38.62          & \textbf{0.157} \\
SSL+Mel                       & SSL+Mel        & 7.60          & 39.68          & 0.118          \\
SSL+Mel                       & SSL+Mel(FT)    & 7.44          & 39.46          & 0.083         
\end{tabular}
\label{tab:my-table}
\end{table}

\subsubsection{Overall ELVC results}
In this section, we compare the performance of CDVAE models and vocoders trained using different feature combinations. We first pre-trained the CDVAE model using Mel as the NL$_{\rm{F1}}$ features and either SP or SSL as the NL$_{\rm{F2}}$ features (see Fig. 1). We then used Mel as the NL$_{\rm{F1}}$ and EL$_{\rm{F1}}$ features to train the ELVC model, which takes Mel as input and generates Mel as output. The Mel features were then fed to the PWG vocoder to generate speech audio. For the first combination, we used Mel+SP for the CDVAE model and Mel for the vocoder. For the second combination, we used Mel+SSL for the CDVAE model and Mel for the vocoder. The results are presented in the first and second rows of Table 3.


We also investigated ELVC systems whose vocoders used the SSL features. In the third row of Table 3, we used SSL+Mel for the CDVAE model and SSL for the vocoder. To further improve the performance, we fine-tuned the vocoder using the converted SSL features obtained by the CDVAE model. The results of this approach are shown in the fourth row of Table 3.

Finally, we evaluated ELVC systems where both the CDVAE model and the vocoder used SSL+Mel features. The results without and with vocoder fine-tuning are presented in the fifth and sixth rows of Table 3, respectively. Note that we had to implement the CDVAE model twice with Mel+SSL and SSL+Mel feature combinations to obtain the converted Mel and SSL features for the vocoder.

Analyzing the results in the first three rows of Table 3, we observe that incorporating the SSL features improves the performance of the ELVC task, especially when using the SSL vocoder to generate the converted audio. Furthermore, we achieved the best overall performance by fine-tuning the vocoder with the converted features, as evidenced by the results in the fourth row of Table 3.

As shown in the fifth and sixth rows of Table 3, the use of cross-domain features (SSL+Mel) in the vocoder did not improve the performance of the ELVC task, which is contrary to the results in Table 2. This result may be attributed to the fact that the Mel and SSL features were generated independently in two separate operations. Consequently, a simple combination of Mel and SSL features without further refinement would not yield better results. Using cross-domain features in the vocoder to improve ELVC performance will be the focus of our future work.


\begin{table}[h]
\caption{Comparison of ELVC systems with SSL-only features and cross-domain features. In both ELVC systems, the vocoder used the SSL features.}
\centering
\begin{tabular}{c|ccc}

Model     & MCD           & F0 RMSE        & F0 CORR    \\   \hline  

VAE(SSL)                            
& 7.39          & \textbf{50.01}          & 0.015       
\\
CDVAE(SSL+Mel)                
& \textbf{7.03}          & 51.21          & \textbf{0.041}

\end{tabular}

\label{tab:my-table}
\end{table}

\subsubsection{Investigating the need for cross-domain features in the ELVC task}
In this section, we examine whether it is necessary to exploit cross-domain features and investigate whether using only SSL features can achieve the best performance in the ELVC task. Since the CDVAE model requires at least two types of input features, we used a variational autoencoder (VAE) model for NLVC pre-training when only the SSL features were used. The other training steps are the same for CDVAE- and VAE-based systems. As shown in Table 4, we observed that better performance using cross-domain features compared to using only the SSL features. The two types of features (Mel and SSL) complemented each other, and the experimental results confirmed that utilizing cross-domain features improved the performance for the ELVC task.

\subsubsection{Subjective listening test}

Finally, listening tests were conducted to further validate the effectiveness of the proposed ELVC system. For intelligibility, the evaluation criteria are as follows: 5 means that every word in the sentence can be understood; 4 means that a few words in the sentence cannot be understood, but it does not affect the understanding of the sentence; 3 means that nearly half of the words in the sentence can be understood, and the content of the sentence can be roughly judged; 2 means that only a few words in the sentence can be understood, but not the whole sentence; and 1 means that the sentence cannot be understood at all. The mean opinion score (MOS) was used to assess speech quality on a scale from 1 to 5, with 1 being the worst and 5 being the best.
From Table 5, Our findings confirm that the CDVAE model with SSL+Mel features achieved better performance in terms of intelligibility compared with using the original SP+Mel features. By fine-tuning the vocoder, we obtained audio signals with higher intelligibility and quality scores.

\section{Conclusions}
In this paper, we proposed a novel ELVC system that uses cross-domain features, that is, a combination of spectral features and SSL representations. We first demonstrated that, by using cross-domain features, a vocoder could be trained to achieve better results for the NLVC task. Next, we confirmed that the cross-domain features could improve the conversion model, leading to better performance for the ELVC task. Finally, we further improved the overall performance by fine-tuning the vocoder to match the output of the conversion model in terms of objective and subjective evaluations. In the future, we will explore the effectiveness of cross-domain features by using different conversion and vocoder model architectures. Moreover, we will investigate the use of SSL features in multimodal ELVC tasks.

\begin{table}[h]
\caption{Subjective evaluation results of ELVC systems with different feature combinations in terms of intelligibility and quality.}
\centering
\begin{tabular}{cc|cc}
CDVAE & Vocoder    & Intelligibility & Quality \\ \hline
SP+Mel           & Mel        & 1.7            & \textbf{3.81}      \\
SSL+Mel        & SSL    & 2.5             & 2.8       \\
SSL+Mel        & SSL(FT) & \textbf{2.6}             & 3.27     
\end{tabular}

\label{tab:my-table}
\end{table}

\bibliographystyle{IEEEtran}

\end{document}